\documentclass{article}
\usepackage[utf8]{inputenc}
\usepackage{makeidx}  % allows for indexgeneration
\usepackage{graphicx}
\usepackage{balance}
\usepackage{listings}
\usepackage{multirow}
\usepackage{threeparttable}
\usepackage{booktabs}
\usepackage{epstopdf}
\usepackage{amsmath,amsfonts,amssymb,mathrsfs,amsthm}
\usepackage{float}                              %per imporre QUI con [H]
\usepackage{multirow}
\usepackage{caption}
\usepackage{subfig}
\usepackage{footnote}
\usepackage{color}
\usepackage{pifont}
\usepackage{acronym}
\usepackage{lscape}

\usepackage{pgfplots}
\pgfplotsset{compat=newest}
\usetikzlibrary{plotmarks}

\newlength\figureheight
\newlength\figurewidth
\definecolor{rosso}{RGB}{220,57,18}
\definecolor{giallo}{RGB}{255,153,0}
\definecolor{blu}{RGB}{102,140,217}
\definecolor{verde}{RGB}{16,150,24}
\definecolor{viola}{RGB}{153,0,153}

\makeatletter

\tikzstyle{chart}=[
    legend label/.style={font={\scriptsize},anchor=west,align=left},
    legend box/.style={rectangle, draw, minimum size=5pt},
    axis/.style={black,semithick,->},
    axis label/.style={anchor=east,font={\tiny}},
]

\tikzstyle{bar chart}=[
    chart,
    bar width/.code={
        \pgfmathparse{##1/2}
        \global\let\bar@w\pgfmathresult
    },
    bar/.style={very thick, draw=white},
    bar label/.style={font={\bf\small},anchor=north},
    bar value/.style={font={\footnotesize}},
    bar width=.75,
]

\tikzstyle{pie chart}=[
    chart,
    slice/.style={line cap=round, line join=round, very thick,draw=white},
    pie title/.style={font={\bf}},
    slice type/.style 2 args={
        ##1/.style={fill=##2},
        values of ##1/.style={}
    }
]

\pgfdeclarelayer{background}
\pgfdeclarelayer{foreground}
\pgfsetlayers{background,main,foreground}

\newcommand{\pie}[3][]{
    \begin{scope}[#1]
    \pgfmathsetmacro{\curA}{90}
    \pgfmathsetmacro{\r}{1}
    \def\c{(0,0)}
    \node[pie title] at (90:1.3) {#2};
    \foreach \v/\s in{#3}{
        \pgfmathsetmacro{\deltaA}{\v/100*360}
        \pgfmathsetmacro{\nextA}{\curA + \deltaA}
        \pgfmathsetmacro{\midA}{(\curA+\nextA)/2}

        \path[slice,\s] \c
            -- +(\curA:\r)
            arc (\curA:\nextA:\r)
            -- cycle;
        \pgfmathsetmacro{\d}{max((\deltaA * -(.5/50) + 1) , .5)}

        \begin{pgfonlayer}{foreground}
        \path \c -- node[pos=\d,pie values,values of \s]{$\v\%$} +(\midA:\r);
        \end{pgfonlayer}

        \global\let\curA\nextA
    }
    \end{scope}
}

\newcommand{\legend}[2][]{
    \begin{scope}[#1]
    \path
        \foreach \n/\s in {#2}
            {
                  ++(0,-10pt) node[\s,legend box] {} +(5pt,0) node[legend label] {\n}
            }
    ;
    \end{scope}
}

\acrodef{PV}{photovoltaic}
\acrodef{HEMS}{home energy management system}
\acrodef{NILM}{non-intrusive load monitoring}
\acrodef{HMM}{Hidden Markov Model}
\acrodef{FHMM}{Fractional Hidden Markov Model}
\acrodef{HSMM}{Hidden Semi-Markov Models}
\acrodef{PF}{particle filtering}
\acrodef{ILM}{intrusive load monitoring}
\acrodef{FSM}{Finite State Machine}
\acrodef{RMSE}{root mean square error}
\acrodef{ACC}{Accuracy}
\acrodef{SPARQL}{SPARQL Protocol and RDF Query Language}
\acrodef{RDF}{Resource Description Framework}
\acrodef{SPIN}{SPARQL Inferencing Notation}
\acrodef{BLH}{battery load hiding}
\acrodef{LLH}{load based load hiding}
\acrodef{OWL}{ontology web language}

\definecolor{orange}{rgb}{1,0.5,0}

\newcommand{\cmark}{\ding{51}}%
\newcommand{\xmark}{\ding{55}}%

\begin{document}



\title{Integration of Legacy Appliances into Home Energy Management Systems}


 \author{
Dominik~Egarter, Andrea~Monacchi, Tamer~Khatib, \\
and~Wilfried~Elmenreich\\[0.3em]
{\small Institute of Networked and Embedded Systems}\\
{\small Alpen-Adria-Universit\"at Klagenfurt, Austria}\\
{\small\{name.surname\}@aau.at}
 }
\maketitle

\begin{abstract}
The progressive installation of renewable energy sources requires the coordination of energy consuming devices.
At consumer level, this coordination can be done by a \ac{HEMS}.
Interoperability issues need to be solved among smart appliances as well as between smart and non-smart, i.e., legacy devices.
We expect current standardization efforts to soon provide technologies to design smart appliances in order to cope with the current interoperability issues.
Nevertheless, common electrical devices affect energy consumption significantly and therefore deserve consideration within energy management applications.
This paper discusses the integration of smart and legacy devices into a generic system architecture and, subsequently, elaborates the requirements and components which are necessary to realize such an architecture including an application of load detection for the identification of running loads and their integration into existing HEM systems.
We assess the feasibility of such an approach with a case study based on a measurement campaign on real households.
We show how the information of detected appliances can be extracted in order to create device profiles allowing for their integration and management within a \ac{HEMS}.
\end{abstract}

\lstset{ %
 % language=Prolog,                % the language of the code
  basicstyle=\ttfamily\scriptsize,         % the size of the fonts that are used for the code
  numbers=left,                   % where to put the line-numbers
  stepnumber=0,                   % the step between two line-numbers. If it's 1, each line
                                  % will be numbered
}

\pagestyle{headings}  % switches on printing of running heads
\section{Introduction}

%General Issue
The transition of our energy system from non-renewable energy to sustainable and renewable energy sources takes place at various levels in the grid, including consumers at household level.
Here, energy might be generated locally (e.g., via a photovoltaic system) or usage can be optimized, reduced or shifted in order to match the energy demand to the current grid situation \cite{Palensky2011}.
Therefore, the usage of devices with significant energy consumption must be coordinated.
Carlson \cite{Carlson2013132} shows that in American households $12$ appliance types are responsible for $80\%$ of household electricity consumption, where white goods such as refrigerator, dishwasher and washing machine demand a higher amount of energy than brown goods (e.g., TV).
%Background
At household level, this demands for a better coordination of energy resources, aiming at lowering overall energy consumption and thus costs.
%means that energy consumption and device usage for white goods must be done in a more coordinated way, which requires some effort.
%On the bright side, this increased coordination and energy awareness leads potentially to lower energy cost, lower energy consumption and, hopefully, to a cleaner planet.
%Problem Statement
To handle the coordination effort for a number of household devices with significant energy consumption, there is a need for a \acf{HEMS} to automatically manage energy resources in building environments.
Networking and operating the components of such a \ac{HEMS} can be difficult, since we have to cope with (i) smart appliances~\cite{elmenreich:wises12} from different vendors, (ii) legacy appliances,
i.e., non-smart, legacy devices without a control or communication interface, and (iii) potential changes in the system due to addition or replacement of devices.

This paper describes an open architecture for the integration of different device types. In particular, we propose a layered model that allows to integrate legacy and smart devices and to represent them with an ontology-based appliance and usage model. By adding a \ac{NILM} mechanism we approach the problem of legacy device integration with a detection approach based on power measurements.
This way, we present a complete and extendable \ac{HEMS} solution providing the basis for energy management applications such as advanced monitoring to increase energy awareness, insert-coin applications to reduce energy consumption, and load shifting approaches to better utilize renewable energy sources or to take advantage of time-based tariffs.

Section~\ref{sec:section2} presents a system architecture describing the different interacting components and mechanisms in a five-layer model, from electric layer up to application layer. Smart and legacy devices are handled by driver components that provide a unified mapping of these devices to the upper layers. Section~\ref{sec:privacy} addresses privacy aspects within this architecture. Section~\ref{sec:managementData} describes data modeling and management within the architecture. The concept for a unified modeling and representation of smart and legacy devices is explained in Section~\ref{sec:legacyintegration}. Section~\ref{sec:legacy} explains how a \ac{NILM} approach allows for the identification of legacy devices. 
Section~\ref{sec:casestudy} describes implementation experiences on the presented approaches. Section~\ref{sec:conclusion} summarizes the main contributions to our \ac{HEMS} architecture and concludes the paper.

\section{The \ac{HEMS} at a glance} \label{sec:section2}
%A \ac{HEMS} is a interaction of different components such as appliances (e.g., smart or non-smart) and monitoring units.
%Therefore, this sections provides an overview of common components in a \ac{HEMS} and introduces a novel \ac{HEMS} architecture offering the possibility to interact with the different components in an interoperable way.
A \ac{HEMS} is a system of computing components that can be employed for optimizing energy resources in building environments.
Typically, a HEMS should be able to collect consumption information of devices, as well as monitoring local production from renewable energy sources (e.g., photovoltaics).
This section provides an overview of common components in a \ac{HEMS} and introduces a novel \ac{HEMS} architecture allowing for the seamless integration of heterogeneous components.
\subsection{Components of a \ac{HEMS}}
Typical building blocks of home energy management systems are:
\begin{itemize}
  \item \textit{Smart meter:}
  The presence of a smart meter results in the possibility to increase the resolution of consumption data.
  On one hand, this offers a feedback mechanisms to residents, who can get a better understanding of energy use in the building environment.
  Furthermore, energy can be provisioned under adaptive prices, which can reflect the energy available in the grid in order to keep the grid balanced.
  On the other hand, utilities can use demand data to improve the pricing and billing mechanism, as well as to monitor the status of the grid and extract valuable knowledge that can improve decision making.
  %For instance, this could enable companies to automatically extract characteristics of inhabitants, so as to build clusters of customers.
  %A comparison of classification methods for households is provided by \cite{beckel2013class} and \cite{1626400}.
  %In \cite{Beckel:2012}, interesting features were selected during interviews with energy providers.
  %The analysis of consumption traces from more than 3000 households showed the possibility to extract features of residents from metering data.
  %Advances in load disaggregation will also give the possibility to extract the power profile of individual loads out of the overall metering data \cite{Zeifman2011, Egarter2013}.
  %Smart metering is therefore raising privacy concerns for the exploitation of consumption information.
  %Solutions like \cite{Yang2012} can reshape the power profile exposed by the household to minimize disclosure of sensitive information.
  %
  \item \textit{Smart appliances:}
  A smart appliance is a device that embeds a computing unit and a network interface, which allows for the interaction with users and other appliances.
  Smart appliances are aware of consumed power, based on local measurement units or built-in profiles \cite{elmenreich:wises12}.
  In order to interoperate with other devices in the network, smart devices need to provide a machine-readable description of their features and properties.
  This way, smart building applications can be realized controlling distributed digital sensors and actuators, which can dynamically join and leave the network.
 To cope with mobility and volatility of nodes, service discovery mechanisms are necessary
  A typical example is given by electric vehicles, which are disconnected for use and eventually reconnected for charging.
  \item \textit{Legacy electrical devices:}
  Although a smart building can be realized from the composition of intelligent devices, building management systems need to consider the presence of non-smart devices.
  A possible solution is to connect sensing units to loads and track their consumption.
  So-called smart outlets and smart plugs form a network of distributed sensing nodes, which normally provide also the possibility to remotely switch loads (on/off).
  Since current market solutions do not support identification of connected loads, any processing of consumption data has to be done at application level.
  \item \textit{Home gateway:}
  A residential gateway is used in building automation systems to bridge the home network to the wide-area network.
  Thus, the gateway represents the connection point between the private network and the internet, and as such, it plays a crucial role in ensuring security and privacy.
  In addition, the gateway is also the point where interconnection and interoperation between heterogeneous technologies takes place.
  Sub-networks using specific technologies, such as automation fieldbuses and Zigbee networks, can be managed from the gateway in order to provide a uniform interface to access resources.
  Beside the integration of networked devices working under different technologies and standards, the gateway should also allow for the integration of legacy devices.
  %Indeed, common electrical devices can not be accessed using a network interface but information of their operation would be of value, especially in a HEMS.
  %
\end{itemize}

\subsection{\ac{HEMS} architecture}\label{sec:architecture}
\begin{figure}[h!]
\centering
\includegraphics[scale=0.32]{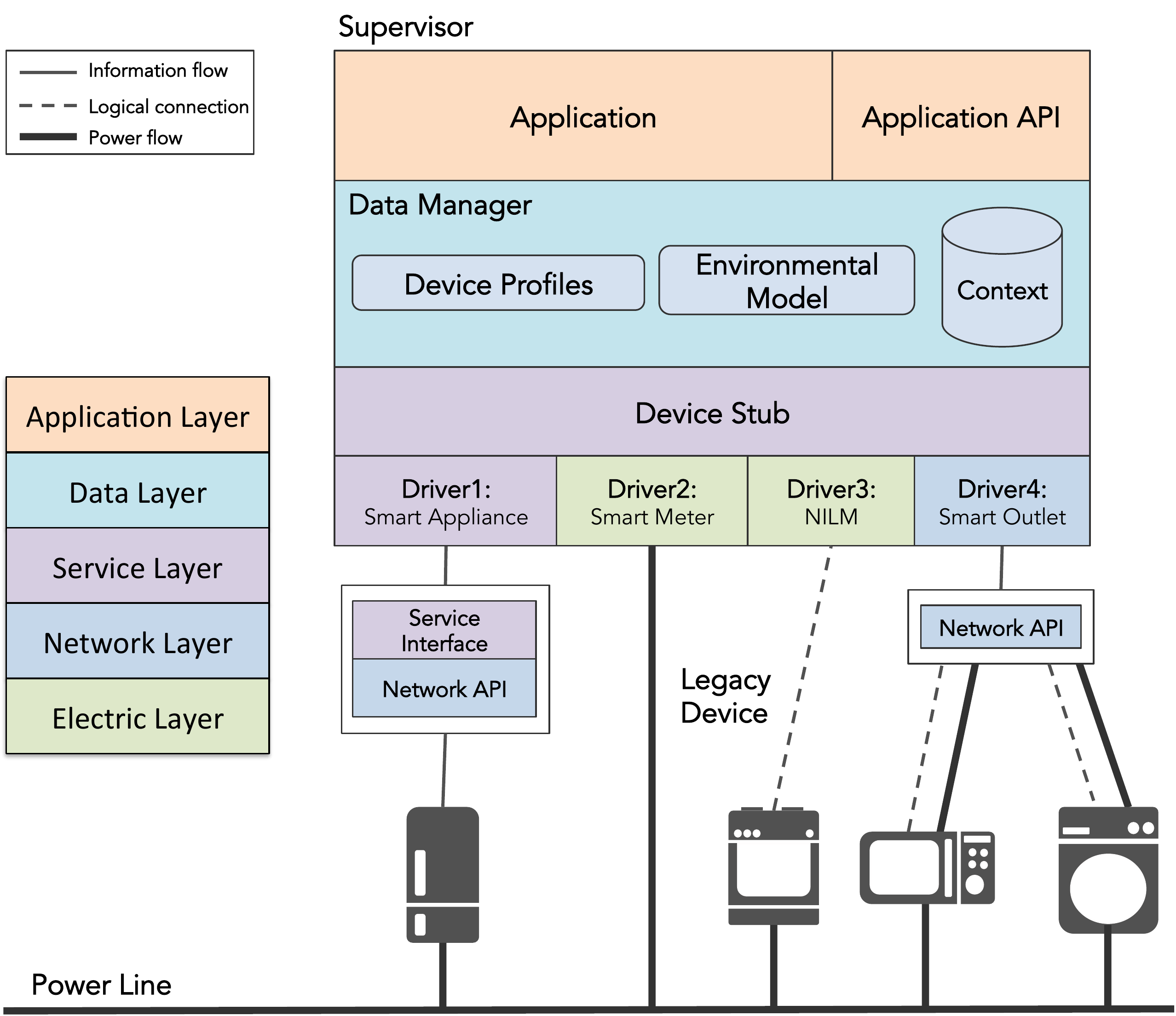}
\caption{Overview of the \ac{HEMS} architecture}
\label{fig:HEMSOverview}
\end{figure}

Figure \ref{fig:HEMSOverview} sketches an architectural model of a \ac{HEMS} that integrates a load disaggregation unit to detect legacy appliances.
In the architecture, specific driver components allows for the detection and management of sub-networks, thus acting as a proxy to integrate networked devices, smart devices and legacy devices.
The model is implemented over the following $5$ layers:
\begin{enumerate}
  \item \textit{Electrical layer:}
  Electrical devices are connected to a common local power distribution network.
  This layer allows devices to deal with electrical power measurements.
  A classic meter works at this level.
   \item \textit{Network layer:}
  This layer provides network connectivity to embedded devices.
  A typical example is given by automation field-buses and wireless sensor networks, such as building automation systems and wireless smart outlets (e.g., ZigBee and WiFi).
  Management of the sub-network requires a specific driver to interface it to the \ac{HEMS}.
  \item \textit{Service layer:}
  In order to be automatically usable by other devices in the network, smart devices are required to provide a machine-readable description of their features and properties.
  The service layer includes the mechanisms by which devices can describe and advertise their features, so that functionalities can be discovered and exploited by other devices \cite{Jammes2005}.
  %The service layer implements the functionalities discussed in Sect. \ref{sec:management}, specifically the description of the device interface and its resources.
  %For instance, a smart appliance can provide a device profile describing its characteristics, as well as exposing functionalities through an interface.
  %Smart environments can be realized controlling distributed digital sensors and actuators, which can dynamically join and leave the network.
  %Service discovery mechanisms are thus necessary to cope with mobility and volatility of nodes. A typical example is given by electric vehicles, which are disconnected for use and eventually reconnected for charging.
  \item \textit{Data layer:}
  The data layer provides an abstract representation of data and functionalities managed by the individual drivers, by providing a homogeneous interface to access this resource.
  This also includes the management of the device profiles, which are datasheets reporting static information of devices (e.g., sensor accuracy, type) and could be stored on the manufacturer's servers.
  In addition, a data model could be employed to perform a basic processing of raw data collected from the drivers, so as to produce a more abstract context representation, which can be stored in a knowledge base.
  %In this way, situational information related to the data coming from the bottom layers can be exploited by context-aware applications, such as decision makers for energy management.
  \item \textit{Application layer:}
  User applications are run in the application layer. For instance a decision maker might rely on the context representation stored in the data layer to react to environment changes.
  A query engine provides an interface between the data and the application layer.
  On the other hand, the network API provides application-level interoperability to the architecture, thus representing a communication interface between applications running on different computing environments.
\end{enumerate}
With exception of the electrical layer, all layers could be implemented locally to the building environment or on remote servers.
For instance, a smart appliance is a device that embeds a computing unit and a network interface.
In order to be integrated in a \ac{HEMS} a smart appliance should implement the first three layers (electrical, network and service),
although the device could scale to the application layer in case data management and decision making at appliance level were necessary.

A \ac{HEMS} is required to deal with distributed resources, built by different manufacturers using different technologies.
As seen, component-level interoperability of networked devices can be achieved by separating the component interface from implemented functionalities.
To achieve such a loose coupling between components in the architecture it is necessary to provide means for:
% Although a smart environment can be obtained from the composition of intelligent devices, current market solutions are still far from being interoperable, especially because there exists no standardized solution to this problem.
%A \ac{HEMS} is required to deal with distributed resources, built by different manufacturers using different technologies.
%Component-level interoperability of networked devices can be achieved by separating the component interface from implemented functionalities.
%In order to be usable within a \ac{HEMS}, smart and legacy appliances are required to provide a machine-readable description of its features, that can be made available to the other devices.
%Architectural reconfigurability can be implemented as composition of loosely-coupled computing components, which separate component interface from implemented functionalities.
%To this end, it is necessary to cope with:
\begin{itemize}
  % Discovery is necessary to keep the resource list update
  \item \textit{Service discovery:}
A service is a self-contained collection of functionalities that is homogeneously provided to users and applications.
Service discovery concerns the discovery and naming of network entities so as to enable resource sharing.
%\cite{Dargie} distinguishes in three generations of name discovery systems, i) \textit{name services} (e.g., DNS) which resolve names to entities and possibly basic attributes, ii) \textit{directory services} (e.g., LDAP) which also provide more complex attribute-based queries so as to retrieve entities satisfying certain attributes, and iii) \textit{service discovery systems}.
%While the first two can work fine in stable environments, such as wired networks, they require configuration and management by human experts.
%On the other hand, service discovery mechanisms can reduce human intervention by adapting automatically to the changing network of devices, thus providing the self-configuration and self-healing properties of autonomic computing \cite{Kephart,Dargie}.
%This aspect is particularly important in mobile wireless networks, where nodes can dynamically join and leave the network.
%To this end, network entities are required to provide description of their functionalities that can be advertised to the other peers.
Generally, service advertisement can be done using broadcast and multicast (e.g., ARP, UPnP), using centralised service registries (e.g., DNS, LDAP, UDDI) or exploiting logical overlays such as with distributed hash tables (e.g., Chord, Kademlia) \cite{Dargie}.
An extensive overview of service discovery mechanisms for embedded systems is provided in \cite{m2membedded}.
  % Interface description in terms of features exposed, classic SOA approach
  \item \textit{Service description:}
%Service description allows for the specification of information related to a service or a device and enables machines to automatically retrieve and use exposed resources.
%Service-oriented architecture (SOA) concerns the organization of computing resources in services, so that applications can developed as loosely coupled compositions of self-contained components.
Providing a simple interface to the service allows for encapsulation of the service complexity and operation.
Services are described in terms of their I/O interface: i) possible operations, ii) constraints on data given and iii) communication protocol.
To ensure format interoperability, service descriptions can rely on formats such as XML or JSON.
%\textcolor{red}{In XML-based webservices a widely used format for messages is the simple object access protocol (SOAP). In addition, the web service description language (WSDL) and the the universal description discovery and integration (UDDI) registry provide respectively an interface definition language and a platform-independent service directory.}

\item \textit{Service coordination:}
%Services are computing components exposing functionalities through a well defined interface.
Complex services can be built by combining individual services.
The two main service composition approaches are service orchestration and service choreography.
In the former, the overall dynamics are defined by an orchestrator component which controls the specific service components.
%\textcolor{red}{In this sense, a standard is the Business Process Execution Language for Web Services (BPEL4WS).}
%The Business Process Execution Language for Web Services (BPEL4WS) represents a standard in this sense.
In the choreography approach the global goal of the system is implemented as local rules in the individual services.
%The behaviour of individual components can be represented as automata and formally analyzed using process algebras (e.g., pi-calculus).
%Coordination of entities might be hard-coded into the service interface, although this might not be applicable in networks of numerous entities and multiple business processes.
%The Web Services Choreography Interface (WSCIspec) and the Web Services Choreography Description Language (WS-CDL) are standardised choreography definition languages.
Choreography definition languages can be used to automatically generate rules for individual services given the overall business process.
%\textcolor{red}{The Web Services Choreography Interface (WSCIspec) and the Web Services Choreography Description Language (WS-CDL) are standardised choreography definition languages.}

\item \textit{Resource description:}
%Beside the description of computing components (e.g., smart appliances) by means of service-oriented technologies it is necessary to identify common data formats and models.
%Indeed, service-oriented technologies tend to offer a black-box model of computing components.
%Therefore, while it provides device interoperability towards the so called plug\&play property\cite{pitzek:05}, data interoperability is still an open issue.
While the plug\&play property \cite{pitzek:05} can be achieved from the description of computing components, data interoperability is still an open issue.
%\textcolor{red}{For the reduction of cost of sensor networks is opening to the so called internet of things (IoT), techniques to cope with data heterogeneity are necessary.}
\cite{Heath2011,6069708,monacchi:2013} %RSLGPID12,}
suggest to semantically annotate sensor data using principles of the semantic web and the linked data initiative.
Beside the Internet of linked documents, the initiative proposes the use of the \ac{RDF} to describe data in terms of relationships.
Specifically, in the RDF data model, the basic information unit is a $\langle$\textit{subject}, \textit{predicate}, \textit{object}$\rangle$ triple.
Data can be related to concepts defined in well-defined vocabularies (i.e., ontologies) so as to be interpretable by all entities sharing the same definition.
The semantic sensor network ontology (SSN-XG) represents a step to this end. Ontologies for data modeling in \ac{HEMS} are presented in Section \ref{subsect:ontologies}.
\end{itemize}

\section{Privacy aspects}\label{sec:privacy}
Privacy concerns arise from handling private information, as it might results in theft and alteration.
Appliances act in a private environment and alteration of consumption and price information in demand response scenarios would entail more than just discomfort on residents activities.
Moreover, smart appliances might be delegated tasks from users to act on their behalf, for instance a smart fridge might be allowed to directly buy goods using a credit card.
Besides, privacy concerns arise from the increasing availability of fine-grained power consumption data in the context of smart metering.
\cite{MolinaMarkham2010} reviews the impact on privacy of different stakeholders, such as the energy utility, marketing/advertisement partners, creditors, the press, and criminals \cite{Skopik}.
Energy consumption data allows for the extraction of usage patterns, describing time of use of devices, and leading to activity recognition and user profiling \cite{Nguyen2013244,Lisovich2010}.
An extreme example of the power of smart metering data analysis was shown in \cite{Greveler2012}.
Therein, the authors have shown the possibility to identify multimedia content, such as the program currently watched on the TV.

Load hiding techniques were presented to alter device operation so as to preserve privacy in households and ensure ownership of energy information.
These techniques can be divided into \ac{BLH} and \ac{LLH} approaches \cite{Egarter2014}.
\ac{BLH} approaches obfuscates metering data using a controllable battery.
The battery is charged and discharged at strategic times to flatten the household’s energy demand.
In contrast, \ac{LLH} uses controllable energy-intensive loads to introduce noise in the daily overall household power draw and obfuscate metering data.
For the purpose, devices used daily and not directly operated by users are favoured in order to minimize the discomfort produced by their altered operation.

\section{Data Management}\label{sec:managementData}
As seen in Sect. \ref{sec:architecture}, \ac{HEMS} can benefit of a data-driven abstraction.
A data manager is intended to provide a homogeneous interface to access static and dynamic data describing the environment (see Fig.\ref{fig:HEMSOverview}).
This requires modeling the environment in which the system operates, using a shared vocabulary to provide semantics.
Static information includes device profiles, reporting characteristics and functionalities, as well as usage models that can be exploited at application level.
Dynamic information represents the environment state.
Sensor measurements are represented abstract context representation that can be utilized by decision makers in a unified way.

\subsection{Data Modeling}\label{subsect:ontologies}% Ontologies for home energy management systems
To achieve data interoperability between data producers and consumers it is necessary to model both data and environment dynamics.
Ontologies can be employed as a unified approach to the semantic annotation of devices, processes, user preferences and activities.
Typically, a \ac{HEMS} requires the specification of the following domains:

\begin{itemize}
  \item \textit{Building information:}
  This includes modeling of the dwelling, such as building geometry and insulation information \cite{CESBP2013_Kofler}.
  \item \textit{Building automation and device description:}
  This includes service orientation \cite{Stavropoulos:2012,Preuveneers04towardsan} and building automation \cite{dogont}.
  \item \textit{User information and preferences}
  This includes living processes in terms of appliance operation and system settings (e.g., for thermal comfort) \cite{EG_ICE2013_Kofler}.
  \item \textit{Energy management:}
  This includes modeling of energy generation, management \cite{Reinisch:2011} and optimization through rule-based reasoning \cite{Tomic:2010}.
  \item \textit{Weather and climate modeling:} This includes modeling of both climate and weather conditions, as well as weather forecasts \cite{IASTED2012_Kofler}.
  \item \textit{Measurement units and sensors:}
  This includes sensor modeling \cite{websem312}, physical phenomena \cite{sweet}, measurement units \cite{QUDT}, as well as geographic information \cite{WGS84}.
\end{itemize}
The use of an ontologies for interpretation of data reuse and integration of existing ontologies to extend the ontological framework to a large scale.

\subsection{Interfaces and Query Languages}
The use of knowledge representation technologies in the semantic web context made tools for the description and retrieval of data widely available.
The \ac{SPARQL} is a query language to handle data within the \ac{RDF} framework, allowing for the addition, deletion, and modification of data triples.
As triples can be added or modified as consequence to the arrival of specific events,
\ac{SPARQL} can also be used to perform complex event processing, by defining rules and constraints over data using the \ac{SPIN}.
While \ac{SPARQL} was designed for static networks, such as the web, data collected in real environments tend to be high dynamic and demand for different query languages able to tackle such a volatily.
Various alternatives have been proposed: C-SPARQL, SPARQLstream, EP-SPARQL, and CQELS.

\section{Integration of Legacy and Smart Appliances}\label{sec:legacyintegration}
In this section we identify the main requirements for the integration of electrical devices within a \ac{HEMS}.
This includes the detection of operating loads and the description of sensed and inferred information using the technologies identified in the previous section.

\subsection{Appliance proxy in the \ac{HEMS}}
%\subsection{Integration in the HEMS}
The component providing a uniform representation of physical devices is the \textit{Device Stub}.
The Stub acts as a proxy for remote devices, which are monitored through \ac{NILM}, sub-metered methods or directly connected through specific network interfaces (e.g., smart appliances).
This allows for keeping a locally mapped representation of remote objects, describing their characteristics and status.
In this way, the information can be provided to the upper data management layer and combined to the environmental data (i.e., context) to be used by applications.
Information of smart devices can be directly retrieved through the network, whereas legacy devices need i) to be detected and distinguished from the smart devices already known and ii) to be described in terms of a device profile (static information) and status.
%Continuous monitoring of legacy devices allows for the extraction of information such as type, properties of operational cycles, expected energy used for certain tasks.
At a service level, the appliance proxy can act as a collection of dummy smart appliances,
by modeling their interface definition using a predefined structure, and making it available within the network.
\subsection{Inferring appliance information from consumption data}\label{subsect:inferring}
Consumption data allows for the extraction of status information of electrical loads, thus giving the possibility to track consumption and build basic profiles.
For instance, a profile could gather data representing operational states of the device, which can be characterized by the energy demanded.
In addition, the profile could also keep track of appliance usage in order to extract a model of inhabitants to be used for energy management applications.

Modern \ac{HEMS} can be equipped with appliance-level sensing units measuring consumption of connected loads.
Another possibility to collect consumption data is provided by the increasing availability of smart meters.
\ac{NILM} is the problem of identifying individual loads from the overall household power draw.
Table \ref{tab:datasheet} shows aspects that can be extracted using \ac{NILM} compared to the ones explicitly provided by smart appliances.
In the table, \ac{ILM} represents the case in which a monitoring unit is connected to each load.
Clearly, the complexity of the \ac{ILM} problem scales to a \ac{NILM} problem when groups of devices (e.g., circuits) are monitored using the same unit.
\begin{table}
 \centering
 \begin{tabular}{c|cccc}
\hline
Parameter & Smart & \ac{NILM}  & \ac{ILM} \\
\hline
ID 				& \cmark & \cmark	& \cmark \\
Type 			& \cmark & $\sim$	& \cmark	\\
Controllable 	& \cmark & \xmark	& $\sim$	\\
Current power 	& \cmark & $\sim$	& \cmark\\
Energy per day 	& \cmark & $\sim$	& \cmark	\\
Appliance usage & \cmark & $\sim$	& \cmark	\\
\hline
\end{tabular}
\caption{Appliance parameters for smart and legacy appliances.}
\label{tab:datasheet}
\end{table}
While a smart appliance can provide a complete description (\cmark), load disaggregation can only infer information related to the operation of appliances, with an uncertainty that depends on the used technology ($\sim$).
Moreover, \ac{NILM} does not allow for appliance control.
While a smart appliance can provide full controllability, a smart outlet/plug (\ac{ILM}) can provide only partial controllability of loads.
As indicated in \cite{Carlson2013132} on American households, white goods are critical for the success of demand response programmes.
In order to tailor this approach to the Central European area, the considered appliances are based on an analysis of Austrian and Italian households~\cite{monacchi:2013Nov}, which identified typical consumption scenarios and formulated conservation strategies tailored to the region.
Based on these data, Table \ref{tab:applianceList} lists important appliances that should be identified and integrated in a HEMS, including smart and legacy devices detected through \ac{NILM} and \ac{ILM} approaches.
Devices are classified according to their controllability and the presence of a user.
Another aspect considered in the table is the extent to which the device should be controlled within energy management applications.
Many legacy devices such as stove, water kettle or microwave are user-driven appliances and therefore, they are not good candidates for load control.
Furthermore, the possibility to switch on/off devices is not sufficient for managing appliances where the starting point has to be scheduled, as smart outlets/plugs cannot pause connected devices.
%for appliance where the time of use can be scheduled, the ability to switch an appliance on or off is not sufficient, because smart outlets/plugs cannot interfere in appliance operation states.
For instance a washing machine can not be controlled via a smart outlet because of its multiple operation states, that is, the device would stay in stop mode or start from the beginning when the power is restored.
%Considering the example of the washing machine, we assume that this device could not be controlled by the smart outlet/plug because of its characteristic of multiple operation states.
%Therefore, we claim that it has to be decided in advance to which extend an appliance should be controllable and if a smart appliance appliance is useful
%\footnote{Smart appliances could control appliance for all operation states.}.
In conclusion, to optimize development and energy costs the design of home energy management systems should consider the necessity of smart appliances.
Selected appliances are the ones responsible for most of demand, as presented in Table \ref{tab:applianceList}.
\begin{table}
 \centering
 \begin{tabular}{c|cccc}
\hline
Type			& Controllable & User-dr. & Tested & Power [W]\\
\hline
Fridge			& \cmark & - & \cmark & $\{8,80,230\}$ \\
Lightning		& \cmark & \cmark & - & - \\
Dishwasher		& \cmark & \cmark & \cmark & $\{1900\}$ \\
Oven			& - & \cmark & - & - \\
Microwave		& - & \cmark & - & - \\
Hob				& - & \cmark & - & - \\
Washing mach.	& \cmark & \cmark & \cmark & $\{190,420,1900\}$ \\
TV				& - & \cmark & \cmark & $\{10,160\}$\\
Computer		& - & \cmark & - & -  \\
Water Kettle	& - & \cmark & \cmark & $\{1750\}$\\
Coffee Mach.	& - & \cmark & \cmark & $\{1280\}$\\
Vacuum Clea.	& - & \cmark & \cmark & $\{1200\}$\\
\hline
\end{tabular}
\caption{Relevant devices to be included into a \ac{HEMS}.}
%It shows, which appliances are controllable, user-driven and if they were considered in the real world test scenario.
%The appliance state specific power values are presented as well.}
\label{tab:applianceList}
\end{table}
\subsection{Management and representation of appliance data}\label{subsec:nilmrepresentation}
The proposed approach considers the management of contextual information, which encompasses both static and dynamic data, such as appliance models and sensor data collected across the physical environment.
A first aspect to consider is the format and locality of such data, in order to provide applications with a unique representation for data access.
Context information can be maintained locally to the building or managed by a context provider acting as trustworthy intermediator to application providers \cite{monacchi:2013}.
As shown in \ref{subsect:ontologies}, ontologies can be used as shared vocabularies to model data and allow for interoperability between different entities.
This means that the agreement of shared vocabularies is fundamental to extend the data management framework to a wide scale.
We are focusing on the integration of legacy and smart electrical devices, and to this end it is necessary to model the following aspects:
\begin{itemize}
  \item \textit{Measurement values:}
  Dynamic data includes time varying information, such as measurement data collected from the physical environment.
  Managing this kind of data requires copying with different requirements.
  First of all, the load disaggregation component deals with time series representing power profiles.
  As the memory required is directly proportional to the sampling frequency, alternative representations such as event-based models should be considered.
  For example, storing edges of the signal, i.e. changes of power exceeding a device-dependent threshold helps reducing the memory consumption to handle time series where the highest frequencies are sparsely present
  However, introducing a threshold constitutes a trade-off between information and memory used, as selecting a large threshold leads to coarse grained data and high information loss which complicates the reconstruction of the signal.
  Consequently, the threshold impacts the effectiveness of \ac{NILM} algorithms and therefore of the whole interoperability process. 
  \item \textit{Appliance profiles:}
  Appliance characteristics are critical for the integration of devices, including both interface description and load disaggregation.
  Manufacturers of smart appliances could provide appliance profiles (i.e., datasheets) describing the device within the communication network.
  Information such as manufacturer, type, energy rating and user controllability can be provided beforehand.
  As an alternative, profiles could be provided by a certified trustworthy entity (e.g., profile provider) or annotated by residents and shared via a common platform

  An appliance profile (see Fig. \ref{fig:taxonomy}) describes the operation of an electrical device, by defining its electrical characteristics.
  Electrical devices operate in the physical environment, offering a physical service.
  A physical service defines the operation of the device for a specific task, such as a certain washing cycle for a washing machine.
  This includes the signature describing the operation of the device, as well as the energy demanded and the current status.
  Here we distinguish between devices with permanent consumption, such as fire alarms, and devices, which can operate in multiple states.
  It is important to remark that device profiles allow for the management of appliances.
  Accordingly, a state is defined in terms of peak active power, tolerance to power variations, duration, and two discomfort factors: i) a delay sensitivity (in seconds) determining the responsiveness of the device and ii) an interruption sensitivity (in seconds) describing the tolerance to interruptions of the state.
  For instance, the start of a coffee machine should not be postponed from its request because of an extremely low delay sensitivity, while its water heating state should not be interrupted because of a low interruption sensitivity.
  The status of a service describes the operation (i.e., on, off, or paused), as well as its progress in terms of start time and elapsed duration.

 Furthermore, smart appliances might expose a virtual service within a network, for instance to retrieve temperature values.
  %The possibility to control a physical service through a virtual service, and consequently a network interface, leads to a smart service.
  A virtual service is hence described as a reference to a machine-readable interface.
  A smart appliance may implement various M2M technologies to provide both physical, virtual and smart services.

  As seen in Section \ref{subsect:inferring}, a load disaggregation unit can extract information of detected appliances in order to build appliance profiles.
  In this way, applications can seamlessy access device information for both smart and legacy devices.
  \begin{figure}[h!]
\centering
\includegraphics[scale=0.36]{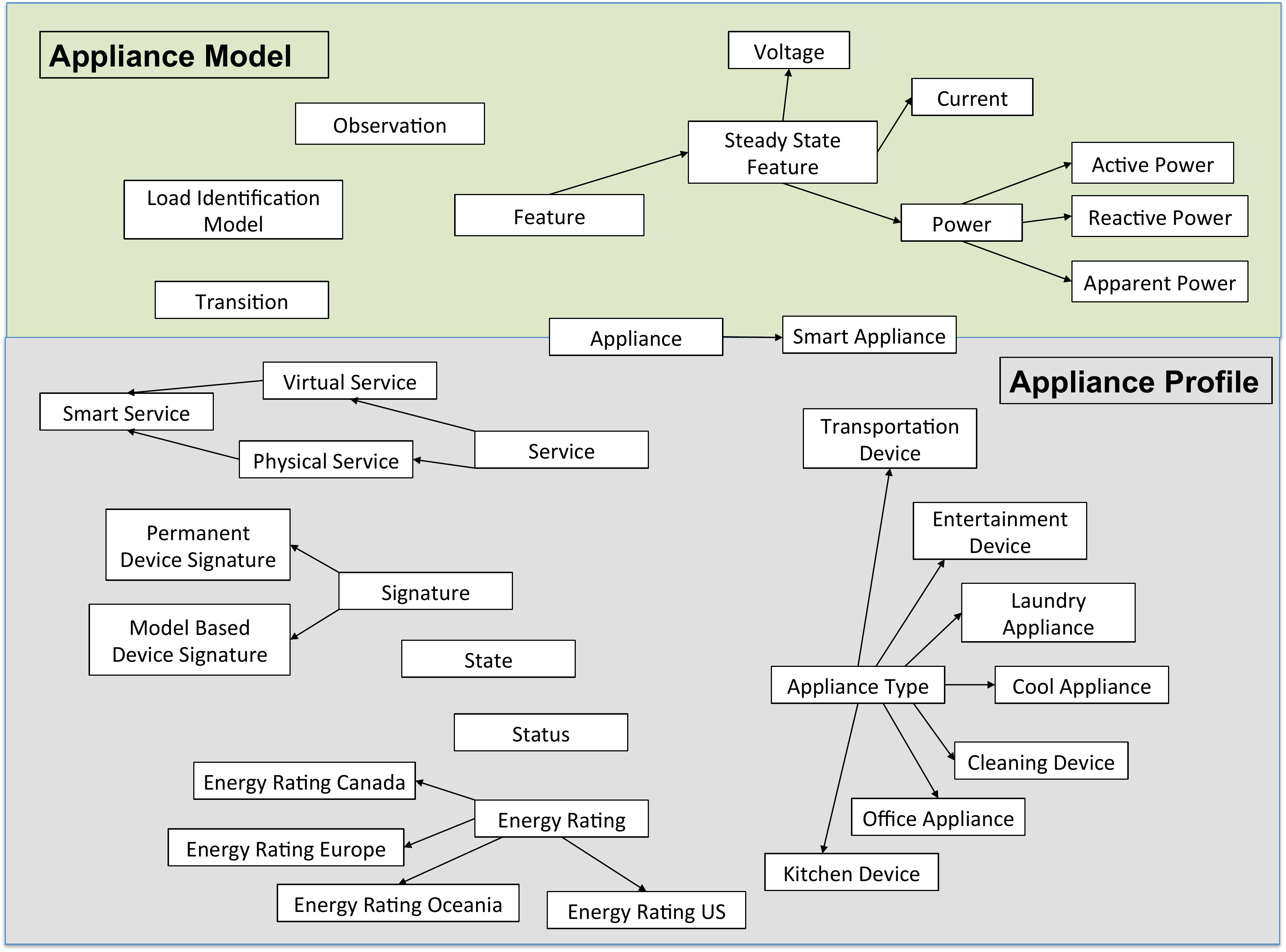}
\caption{Taxonomy of appliance description and model}
\label{fig:taxonomy}
\end{figure}
  \item \textit{Appliance identification models:}
  In order to detect running devices, the load identification component needs to deal with appliance models describing the behavior of devices through a set of observable states.
  %Power values for each operational task could be provided by a third party repository, as well by manufacturers of smart appliances.
  For instance, an appliance model can be defined using state-based representations such as \ac{FSM} and \ac{HMM}, which model the appliance dynamics as a trajectory of state transitions over time.
  Specifically, each device observation is described by a set of features, such as current and power, as well as outgoing transitions to other observable states.
  In addition, certain models can also express the typical duration of a device observation (e.g., \ac{HSMM}).
  To model device dynamics, transitions connecting two observations are generally associated to a transition probability.
  %The used features of the appliance model can be seen as steady state feature or transient feature.
  Fig.~\ref{fig:ontology} reports the ontology for electrical appliances, showing both the device profile and the load identification model.
	  \begin{figure}[h!]
\centering
\includegraphics[scale=0.36]{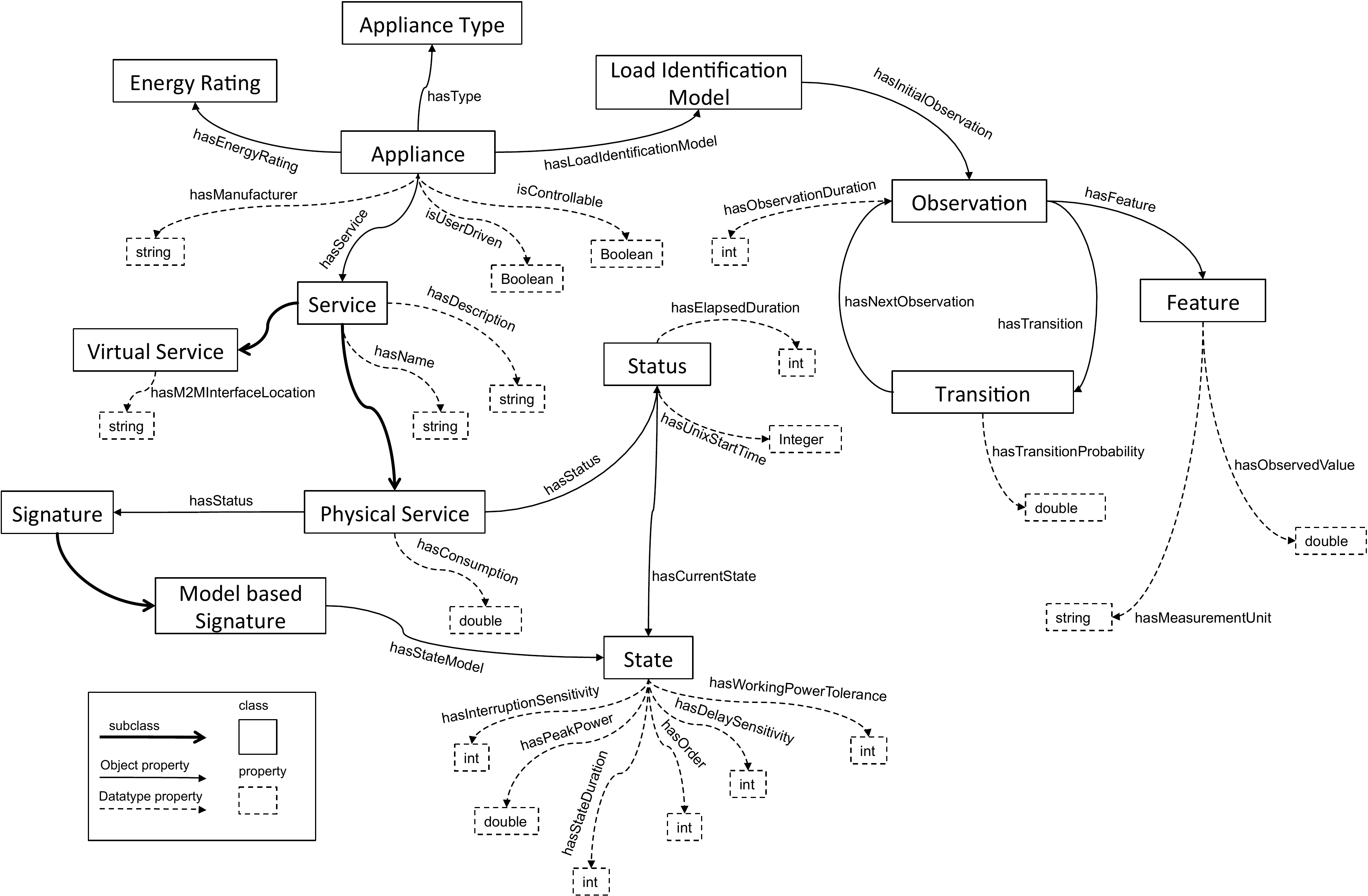}
\caption{Ontology of appliance description and model}
\label{fig:ontology}
\end{figure}
  \item \textit{Appliance usage models:}
  The problem of mining appliance usage patterns concerns the extraction of models of how appliances are used by residents, given a sequence describing changes on their operational state.
  Given logged data describing status changes of appliances, representing relationships between the usage of different devices can be solved by data mining algorithms. % such as association rule mining, episode-generating \ac{HMM}s \cite{eps351238} and Bayesian network learning \cite{ploix}.
  In \cite{Andrea2014}, we apply Bayesian network to learn the usage model of a coffee machine based on real consumption data.
  As models strictly depend on end-user applications, their construction and management needs to be done at application level.
  The details of applying data mining for generating appliance usage models is beyond the scope of this paper.
\end{itemize}
\section{Appliance detection and classification}\label{sec:legacy}
To monitor and integrate legacy devices into a \ac{HEMS}, we can distinguish three monitoring and detection possibilities: a single meter approach (\ac{NILM}),
a multiple meter approach at device level (\ac{ILM}) and a multiple meter approach for device groups.

\subsection{Non-intrusive load monitoring - \ac{NILM}}\label{subsec:nilm}
\ac{NILM} is the problem of disaggregating running loads from overall consumption data.
\ac{NILM} is a single-meter approach.
The approach was first introduced by Hart \cite{Hart1992}.
The method detects running appliances according to specific characteristics of their power signature.
%The NILM approach is single-sensor based and can be used to integrate legacy appliances because of its ability to detect white goods dependent on the used NILM approach.
%The increasing resolution of power measurements as a consequence of the smart meter roll-out is going to provide a means for the detection of appliances.
%Appliances that can be identified by power profile characteristics can be identified from the total power profile collected by the meter.
State-of-the-art \ac{NILM} algorithms can be distinguished into supervised and unsupervised approaches.
Supervised techniques are based on labeled data and can be divided into optimization and pattern recognition based algorithms.
In the latter case, the \ac{NILM} problem is considered as an optimization problem: a total power consumption and a database of known power profiles of appliances are given.
A composition of devices' power profiles are chosen from the database to approximate the total power consumption with minimal error \cite{Liang2010,Egarter2013,Suzuki2008}.
Pattern recognition techniques include clustering approaches \cite{Hart1992}, Bayesian approaches  \cite{Zeifman2012}, neural networks algorithms \cite{Srinivasan2006} and support vector machines \cite{Srinivasan2006,Lin2010}.
The disadvantage of supervised learning and classification is the need of labeled data during a training phase, which implies greater development costs and effort.
Lately, recent research in load disaggregation is focusing on unsupervised algorithms.
Unsupervised algorithms are not using any labeled data and consequently, no training phase.
Recent \ac{NILM} algorithms are based on k-means clustering \cite{goncalves_unsupervised_2011}, \ac{FHMM}, and its variants \cite{zico2012,Zaidi2010,Kim2011,Zoha2013}.
State-of-the-art \ac{NILM} approaches are highly dependent on the used sampling frequency and have several limitations.
According to \cite{CarrieArmel2013}, it is possible to detect approximately $10$ different appliances with a sampling frequency of seconds.
The higher the sampling frequency, the more meaningful are the device features and the more accurate is the appliance classification.
Currently, there exists no load disaggregation technology which can solve the problem in all its aspects. Open challenges and limitations are:%
%Up to now no NILM approach is working perfect and can solve the problem of load disaggregation in all its diversity.
% Therefore, we list in the following possible issues and limitations of NILM approaches to solve the problem of disaggregated devices:
\begin{enumerate}
  \item Noise interference: presence of noise, which is typical to a measurement process,  affects the quality of extracted device features;
  \item Appliance modelling: differences in type, characteristics and manufacturer make the modeling of devices more complex. This means that a model of a washing machine does not allow a load disaggregation algorithm to correctly detect another type of washing machine with different characteristics.
  \item The quality of extracted features is directly proportional to the sampling frequency \cite{CarrieArmel2013}.  %Table \ref{tab:NILM_Limitation} reports features and number of detectable appliance according the chosen sampling frequency. %the used sampling frequency against usable features and detectable features are listed.	
  \item Similar appliance features: different appliance types can behave in a similar way which makes it difficult to distinguish between them.
  Therefore, more advanced appliance characteristic features are needed.
  \item Unknown appliances in the power draw add uncertainty to the disaggregation process.
  \item Degree of overlapping power draws: concurrently operating devices imply overlapping power profiles. Therefore, their number determines the complexity of the classification process.
   %the more appliances are used, the higher is the probability that appliance usages are overlapping and therefore, disturbing the classification process by itself.
  \item Unpredictable appliance usage: user behaviour introduces further uncertainty to the disaggregation process. Resident's habits, such as time of use and typical duration, tend to be user specific.
  %time and time duration of appliances usage are different from consumer to consumer. Each consumer has its own habits to use each appliance.
  \item Unknown number of appliances: the number and types of appliances used vary from home to home.
  \item Computational complexity: Each approach has different computational and complexity costs, which have to be considered for the specific application purpose \cite{MakoninPG13}.
\end{enumerate}
A disaggregation technique should be selected considering both application and hardware costs, as depending on the selected technique, different limitations and restrictions arise. %%can be dealed effectively or not.
Zeifman suggests in \cite{Zeifman2012} that a load disaggregation solution should fulfill the following requirements:
\begin{itemize}
  \item Power measurements with $1$ $Hz$ sampling frequency;
  \item The minimum acceptable accuracy is $80$ to $90\%$;
  \item No training should be necessary;
  \item Real-time capability should be provided;
  \item $20$ to $30$ appliances should be detectable;
  \item It should work with different appliance types, such as on/off appliances, multi-state appliances, continuous appliances, as well as permanently operating ones \cite{Zeifman2011}.
\end{itemize}

\subsection{Multiple-meter load monitoring}
The sub-metered approach is intrusive (\ac{ILM}) as it requires the connection to each device of a monitoring unit, such as a smart outlet/smart plug.
Device identification can be performed both by humans, who can specify the type of each connected device, as well as inferred from collected data using the techniques presented in Section \ref{subsec:nilm}.
The presence of device-level consumption information facilitates the appliance classification problem, as the search space is reduced to fewer connected devices \cite{reinhardt12tracebase,Englert2013}.
This allows for a hierarchical load disaggregation, where results of load disaggregation on appliance groups monitored with \ac{ILM} can be combined to obtain the overall solution to device classification \cite{Bergman2011}.
The disadvantage of using multiple meters is the resulting cost for the sensing units, which add an overhead for the collection of the data, as well as maintenance costs.

\subsection{Extracting load identification models}
The load identification component requires models describing the operation dynamics of electrical devices.
Models can be obtained based on {\em a priori} information or can be extracted online from measurements.
An example for building appliance models was firstly presented in \cite{Hart1992}.
This approach clusters appliance events according to active and reactive power to build a finite state machine of the appliances.
Another approach was presented by Dong \cite{Dong2013}. Major residential load signatures are extracted and can be used for \ac{NILM} and load condition monitoring.
Beside extracting load signatures and building appliance models, Parson \cite{parson2012} uses a general appliance knowledge to disaggregate appliances from the metering data.
Recent research on appliance modeling is based on probabilistic graphical models, as they are able to capture both appliance bahaviors, structures and features (e.g.: on/off appliances or multi-stat appliances) as well as user's behavior expressed in a probabilistic way.
The \ac{HMM} is commonly used to model appliances, while variants are employed to incorporate time and other elements.
For example \cite{Wong2014} uses a semi-\ac{HMM} to represent more realistic appliance usage models, whose transitions are not geometrically distributed.

\section{Implementation experiences}\label{sec:casestudy}

\subsection{Particle Filter based Load Disaggregator}
\begin{figure}[h!]
\centering
\includegraphics[scale=0.27]{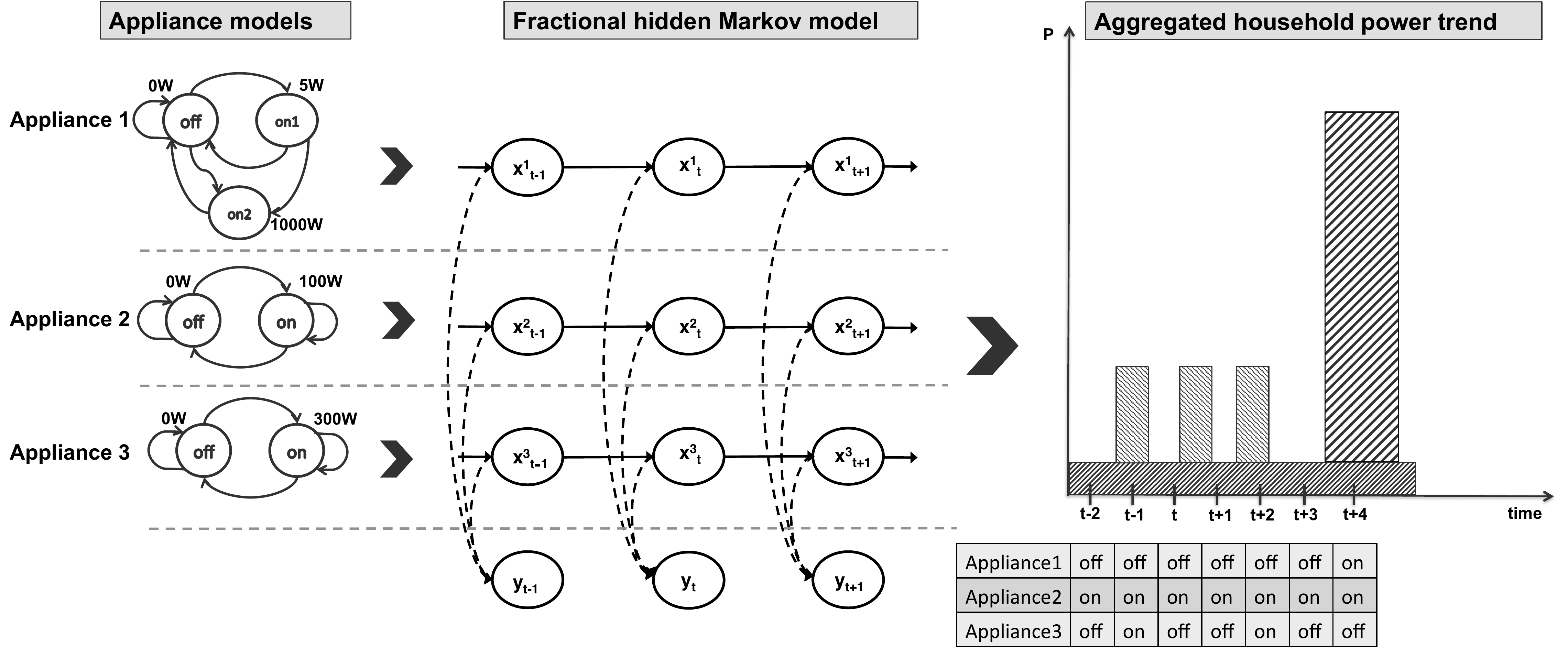}
\caption{Sketch of 3 appliances modeled by \ac{HMM} and representing the aggregated household demand by \ac{FHMM}}
\label{fig:HMM}
\end{figure}

To meet the requirements for a load disaggregation component (Section \ref{sec:legacy}) we used an online approach based on \ac{PF}, as in \cite{Egarter2013:BuildSys}.
Each appliance is modeled by a \ac{HMM} whereas the model is described by its observations and transition behavior between states.
The \ac{HMM} observations are represented by the power demand for each appliance operation state and state transition behavior is described by the operation structure (e.g. on/off appliance with 2 states, multi-state appliance with several operation states) of the appliance.
Accordingly, the \ac{HMM} models the power demand on appliance level.
To model the household power demand, we use the \ac{FHMM} to aggregate each appliance power trend modeled by the \ac{HMM}.
Figure \ref{fig:HMM} sketches the \ac{HMM} modelled appliances, the \ac{FHMM} and the resulting aggregated power demand.

The \ac{PF} aims to approximate the posterior density of the \ac{FHMM} to disaggregate each appliance power demand and appliance state from the household demand, according to the current observed consumption and the given appliance models.
Therefore, the \ac{PF} output estimates the household consumption, whereas a simple decision maker based on thresholding and with knowledge of each appliance model decides for each appliance state.
The use of \ac{PF} as load disaggregator is beneficial for three reasons.
First, \ac{PF} can handle non-linear problems presented by non-linear behaving loads such as a driller or a dimmer.
Second, it can handle non-Gaussian noise influences resulting from uncertainty in power trends and consumption data.
Third, \ac{PF} and its performance can be adjusted by the number of used particles.
The more particles the \ac{PF} considers, the better the estimated posterior density.
Nevertheless, we want to point that the number of particles can not be chosen arbitrarily due to the computational effort of the approximation process.
We empirically identified 1000 particles as an appropriate number balancing the trade-off between the context of computational effort and detection performance for a typical household with $\sim20$ relevant devices.
In the following evaluations, on/off and multi-state appliances are considered whereas the \ac{PF} has the knowledge of the model structure (two or more operating states) and the expected power demand as observation for each state.
Knowledge of the transition matrix values is not necessary since the \ac{PF} is independently estimating the appliance states with an appropriate number of used particles.

\subsubsection{Dataset}
\begin{figure}[h!]
\centering
\includegraphics[scale=0.3]{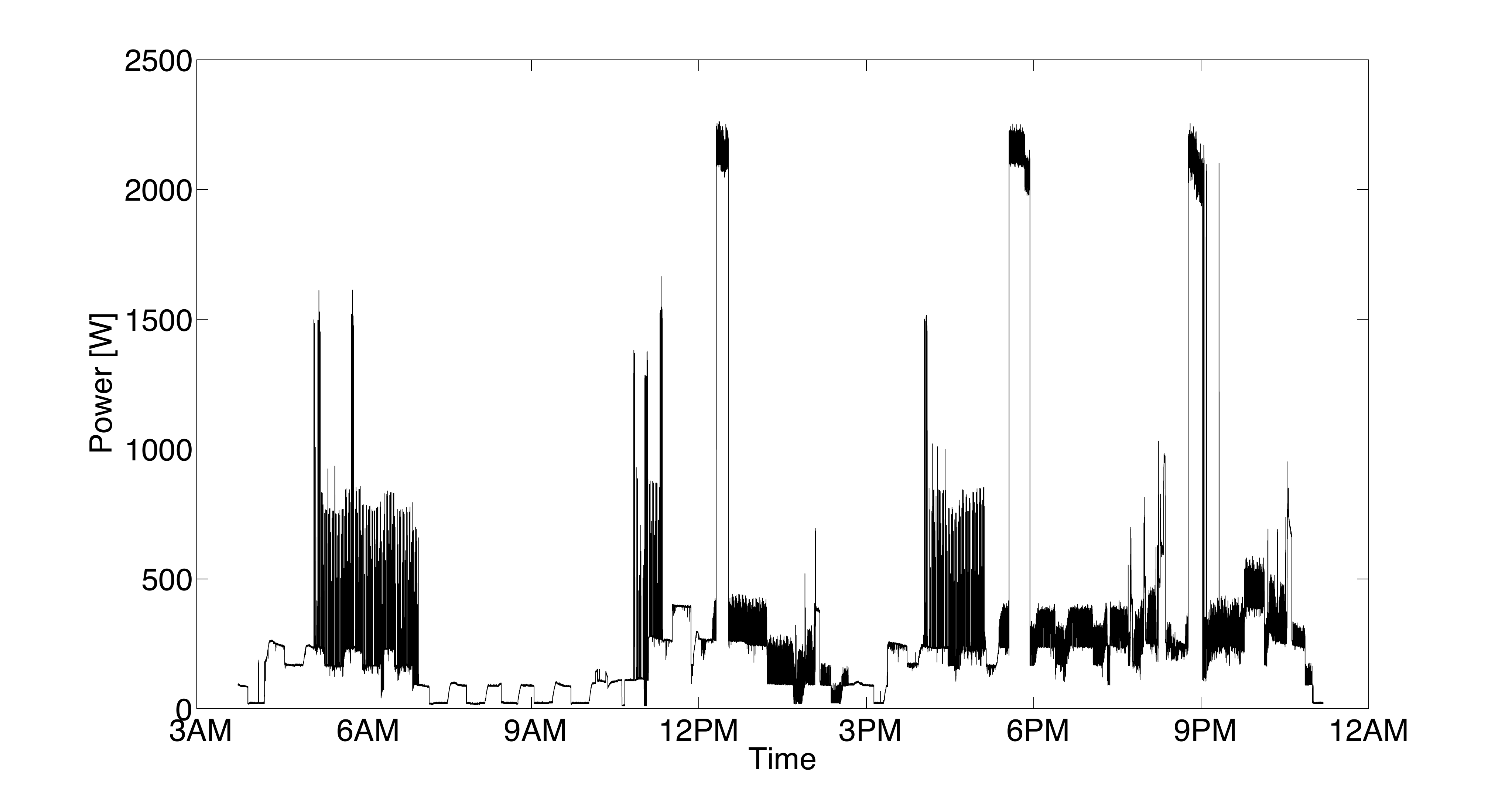}
\caption{Plot of the power consumption trend for seven appliances over several hours}
\label{fig:total}
\end{figure}

An appropriate power consumption dataset has to be used to evaluate our test scenario.
There exists several publicly available datasets, for example the REDD dataset \cite{kolter-kdd-2011}, the AMPds dataset \cite{makonin2013ampds}, and the Smart* dataset \cite{barker2012smart}.
In order to address power consumption in Austria and Italy, we use the GREEND dataset \cite{Andrea2014} containing appliance level power measurements of Austrian and Italian households.
The dataset offers appliance-level active power measurements at 1Hz resolution.
In total 9 houses with 9 typical household appliance were monitored, although our evaluations relies on 7 appliances in house 0 (Table \ref{tab:applianceList}).
We have chosen the devices according to their contribution to the household power demand as in \cite{Carlson2013132}.
Figure \ref{fig:total} presents a plot of the aggregated power demand of seven appliances.

\subsubsection{Test Scenarios}
To show how the selected load disaggregator could be used to integrate legacy appliances we defined three different test scenarios:
\begin{enumerate}
  \item \textit{Disaggregation of overall household (\ac{NILM})}.
  To simulate an approximated overall household demand, the power consumption for each appliance listed in Table \ref{tab:applianceList} is aggregated over time.
  The load disaggregator aims to disaggregate the appliance level consumption data based on predefined appliances models (\ac{HMM}). 
  This test scenario should depict the common household case where one energy meter with corresponding aggregated consumption data is available.
  \item \textit{Appliance groups.}
  It is conceivable that more than one energy meter in form of smart plugs/ smart sockets are available in a household.
  These meters are monitoring a group of connected appliances.
  Therefore, the energy demand of appliances are monitored in groups resulting in a reduction of the appliance number aggregating their power demand.
 The complexity of the disaggregation problem is decreased.
  In order to confirm this assumption, we arbitrarily choose two groups of four and three appliances, aggregate their consumption data and give this information with the appliance models as input to the load disaggregator.
%  \item \textit{Individual devices.}
%  The simplest disaggregation problem is the case where the consumption data is caused by only one appliance out of the group of appliances in Table \ref{tab:applianceList}.
%  Thus, the load disaggregator has the knowledge of appliance models and should detect out of all possible appliance which device is the consuming one.
%   In this scenario, we assume that each appliance in Table \ref{tab:applianceList} is equipped with a smart plug and the load disaggregator aims with the knowledge of the appliance models to detect which appliance is connected to it.
\end{enumerate}
For all three test scenarios, 7 consecutive days with $1$ second measurement granularity are considered.
To make statements how the load disaggregator is performing for the different test scenarios, we are defining the following commonly used evaluation metrics:
\begin{itemize}
  \item \textit{\ac{ACC}.} The accuracy is defined as the number of correctly detected events for the case when a device is on (TN$\ldots$true positive) or off (TN$\ldots$true negative) over all possible events and is defined as:
  \begin{equation}
ACC = \frac{TP+TN}{N} \in [0,1],
\end{equation}
where $N$ represent the number of all consumption samples.
  \item \textit{Normalized \ac{RMSE}.} The \ac{RMSE} is the error between real and estimated power signal normalized by the difference between minimum and maximum power in the signal. The \ac{RMSE} is formulated as:
  \begin{equation}
{RMSE} = \frac{\sqrt{E((\hat{\Theta}-\Theta))^2}}{max(\Theta)-min(\Theta)},
\end{equation}
where $\Theta$ represents the true total power load, $\hat{\Theta}$ the estimated total power load produced by \ac{PF} and $max(\Theta)$ and $min(\Theta)$ the maximum and minimum power value of the total power load.
  \item \textit{Energy disaggregation error} The percentage error of the estimated energy with respect to the real consumed energy, over the duration of the experiment.
\end{itemize}

\subsubsection{Results}

As described before, we considered different test scenarios for our evaluations.
The first one deals with the aggregated power demand of seven appliances listed in Table \ref{tab:applianceList}.
In Table \ref{tab:aggregatedResults} the evaluation results are presented.
The reached \ac{ACC}, \ac{RMSE} and disaggregation error depend on the complexity of devices (e.g.: on/off or multi-state appliances) and the similarity of devices regarding their power demand.
For example, the set of TV, fridge and washing machine yields a decreased \ac{ACC} and \ac{RMSE} due to their appliance type and their similarity of consuming power.

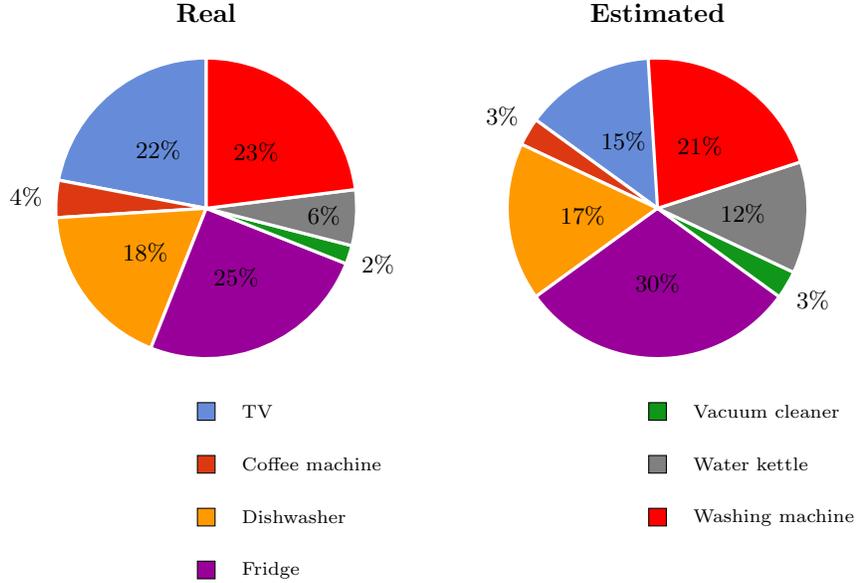
\begin{figure}
\begin{tikzpicture}
[
    pie chart,
    slice type={TV}{blu},
    slice type={coffee}{rosso},
    slice type={dishwasher}{giallo},
    slice type={fridge}{viola},
    slice type={vaccuum}{verde},
    slice type={waterkettle}{gray},
    slice type={washingmachine}{red},
    pie values/.style={font={\small}},
    scale=2
]

    \pie[values of coffee/.style={pos=1.2},values of vaccuum/.style={pos=1.2}]
    {Real}{22/TV,4/coffee,18/dishwasher,25/fridge,2/vaccuum,6/waterkettle,23/washingmachine}
    \pie[xshift=3cm,values of coffee/.style={pos=1.2},values of vaccuum/.style={pos=1.2}]%
        {Estimated}{15/TV,3/coffee,17/dishwasher,30/fridge,3/vaccuum,12/waterkettle,21/washingmachine}

    \legend[shift={(0cm,-1cm)}]{{TV}/TV, {Coffee machine}/coffee, {Dishwasher}/dishwasher,{Fridge}/fridge}
    \legend[shift={(3cm,-1cm)}]{{Vacuum cleaner}/vaccuum,{Water kettle}/waterkettle,{Washing machine}/washingmachine}

\end{tikzpicture}
\caption{Sketch of the household energy partition on appliance level for the measured and estimated case are shown}
\label{fig:aggregatedEnergy}
\end{figure}
In Figure \ref{fig:aggregatedEnergy} the energy partition on the real and the estimated energy computation on appliance level is shown. 
The estimated energy values and their corresponding percentages on the overall energy demand show again the dependence of the load disaggregator on the appliance type and the similarity between appliances.

\begin{figure}
\begin{tikzpicture}
[
    pie chart,
    slice type={TV}{blu},
    slice type={vaccuum}{verde},
    slice type={washingmachine}{red},
    pie values/.style={font={\small}},
    scale=2
]

    \pie[values of vaccuum/.style={pos=1.2}]
    {Real}{47/TV,3/vaccuum,50/washingmachine}
    \pie[xshift=3cm,values of coffee/.style={pos=1.2},values of vaccuum/.style={pos=1.2}]%
        {Estimated}{43/TV,4/vaccuum,53/washingmachine}

    \legend[shift={(-0.5cm,-1cm)}]{{TV}/TV}
    \legend[shift={(1cm,-1cm)}]{ {Vaccuum cleaner}/vaccuum}
    \legend[shift={(3cm,-1cm)}]{{Washing machine}/washingmachine}
\end{tikzpicture}
\caption{Sketch of the household energy partition on appliance level for the measured and estimated case are shown in which appliances are formed into group 0 (TV, vacuum cleaner, washing machine)}
\label{fig:grouped_1_aggregatedEnergy}
\end{figure}
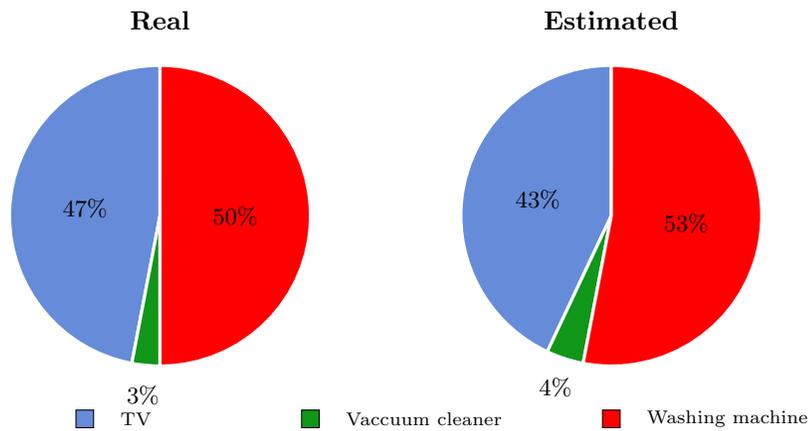

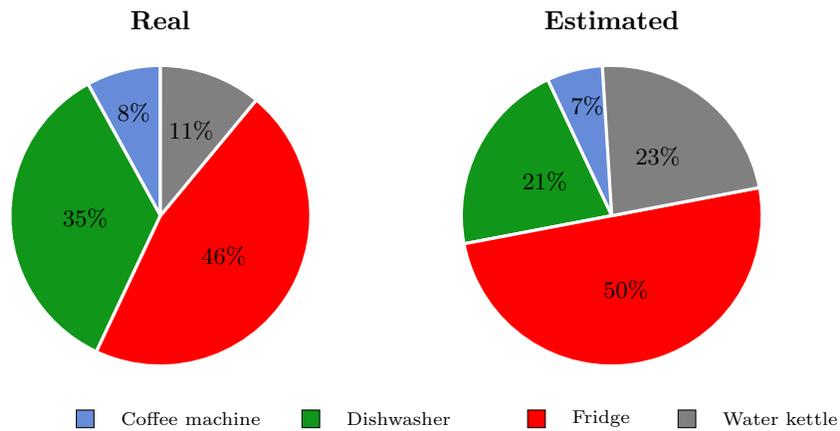
\begin{figure}
\begin{tikzpicture}
[
    pie chart,
    slice type={coffee}{blu},
    slice type={dishwasher}{verde},
    slice type={fridge}{red},
    slice type={waterkettle}{gray},
    pie values/.style={font={\small}},
    scale=2
]

    \pie
    {Real}{8/coffee,35/dishwasher,46/fridge,11/waterkettle}
    \pie[xshift=3cm]%
        {Estimated}{7/coffee,21/dishwasher,50/fridge,23/waterkettle}

    \legend[shift={(-0.5cm,-1cm)}]{{Coffee machine}/coffee}
    \legend[shift={(1cm,-1cm)}]{ {Dishwasher}/dishwasher}
    \legend[shift={(2.5cm,-1cm)}]{{Fridge}/fridge}
      \legend[shift={(3.5cm,-1cm)}]{{Water kettle}/waterkettle}
\end{tikzpicture}
\caption{A Sketch of the household energy partition on appliance level for the measured and estimated case are shown in which appliances are formed into group 1 (coffee machine, dishwasher, fridge, water kettle)}
\label{fig:grouped_2_aggregatedEnergy}
\end{figure}
In the second test scenario appliances are grouped to decrease the number of appliances.
The results of \ac{ACC} and \ac{RMSE} are shown in Table \ref{tab:groupedResults} and are improved compared to Table \ref{tab:aggregatedResults}.
As reason we assume the decreased number of appliances and the corresponding decreased probability to have similar appliances in the same appliance group.
Figure \ref{fig:grouped_1_aggregatedEnergy} and \ref{fig:grouped_2_aggregatedEnergy} show the energy partitions of the submetered demand for the real and the estimated energy values.
As in the previous test scenario, similar appliances influence the result of the disaggregator.
In summary, a load disaggregator could be used to disaggregate appliances from the aggregated household power demand.
The performance is mainly affected by the number of considered appliances, the type of appliances and their similarity.
Note that the presented load disaggregator is aware of the appliance type/model and the number of appliances.
Although it is one of numerous \ac{NILM} approaches applicable for this problem, the presented approach appears to be a good choice fulfilling the requirements given by Zeifman.
Nevertheless, to achieve full integration of legacy devices in a \ac{HEMS}, a load disaggregator has to combine appliance detection with signature extraction approaches.
While appliance detection provides information of appliance status and allows for the inference of device profiles (see Sect. \ref{subsec:nilmrepresentation}),
signature extraction allows for the inference of device models, to be used for improving the appliance detection process.

\begin{table}
 \centering
 \begin{tabular}{|c|cc|}
\hline
&  \ac{ACC} & \ac{RMSE} \\
\hline
\hline
TV 							& 0.8705	& 0.2995 \\
Coffee machine 				& 0.9901	& 0.0673 	 \\
Dishwasher 					& 0.9549	& 0.1263 \\
Fridge						& 0.8992	& 0.2095 \\
Hoover						& 0.9952	& 0.0649 \\
Water kettle  				& 0.9827	& 0.1131\\
Washing machine				& 0.8826	& 0.1251\\
\hline
\hline
Total						& 0.9393	& 0.051 \\
\hline
\end{tabular}
\caption{Results for test scenario 1}
\label{tab:aggregatedResults}
\end{table}

\begin{table}
 \centering
 \begin{tabular}{|c|cc|}
\hline

&  \ac{ACC} & \ac{RMSE}  \\
\hline
\hline
& \multicolumn{2}{c|}{Group 1} \\
\hline
Coffee machine 			& 0.994		& 0.0332   \\
Dishwasher 				& 0.9626	& 0.0941  \\
Fridge 					& 0.994 	& 0,0742 \\
Water kettle			& 0.9889	& 0.0948 \\
\hline
Total 					& 0.9849	& 0.0091  \\
\hline
\hline
& \multicolumn{2}{c|}{Group 2} \\
\hline
Vacuum cleaner			& 0.9643	& 0.1548  \\
Washing machine  		& 0.997		& 0.152	\\
TV 						& 0.9456	& 0.0186  \\
\hline
\hline
Total 					& 0.9699 & 0.0101  \\
\hline
\end{tabular}
\caption{Results for test scenario 2}
\label{tab:groupedResults}
\end{table}

\subsection{Annotation of inferred information}
\begin{figure}[h!]
\centering
\includegraphics[width=\columnwidth]{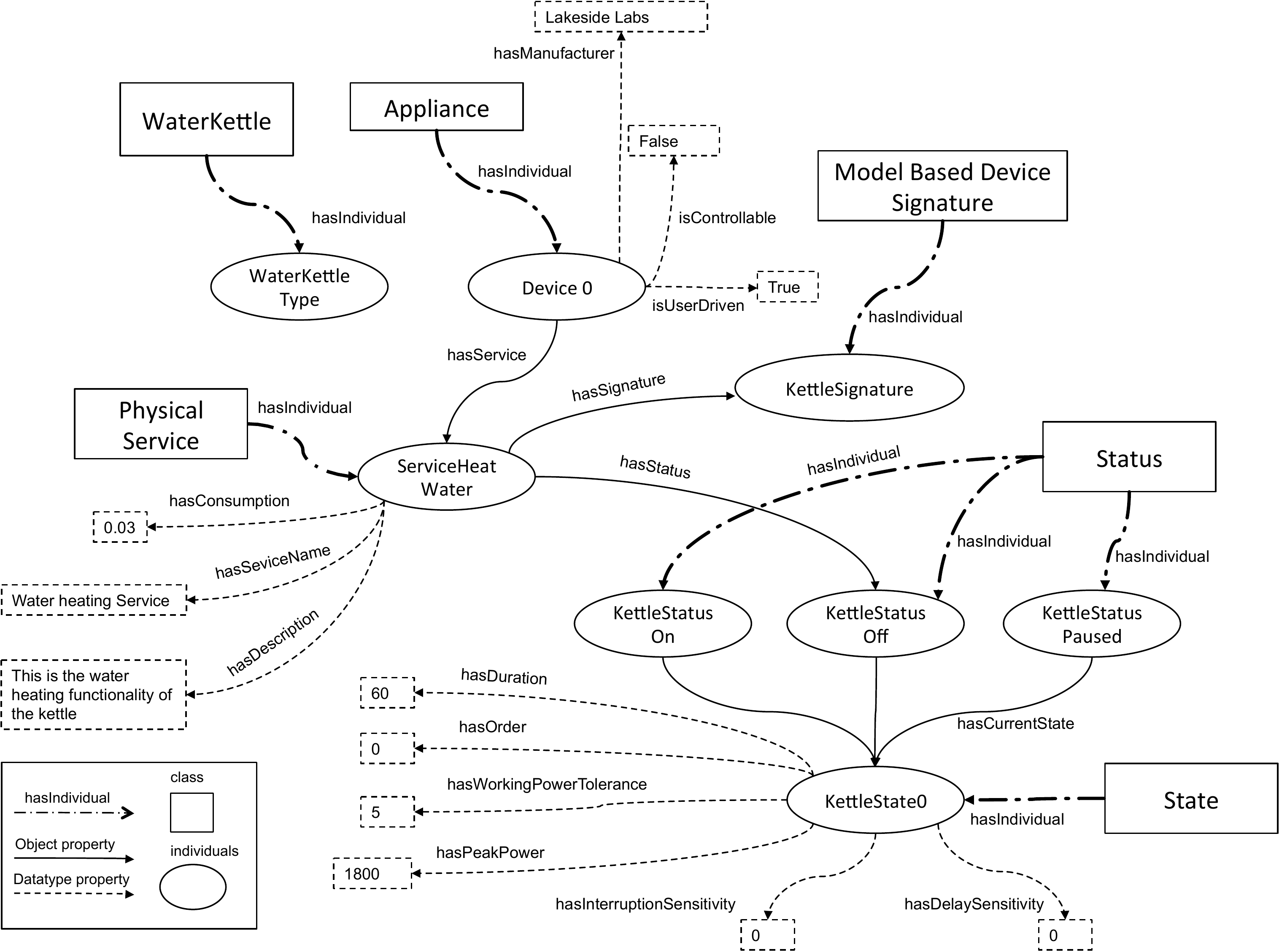}
\caption{Device profile for the water kettle}
\label{fig:datasheetkettle}
\end{figure}

\begin{figure}[h!]
\centering
\includegraphics[width=\columnwidth]{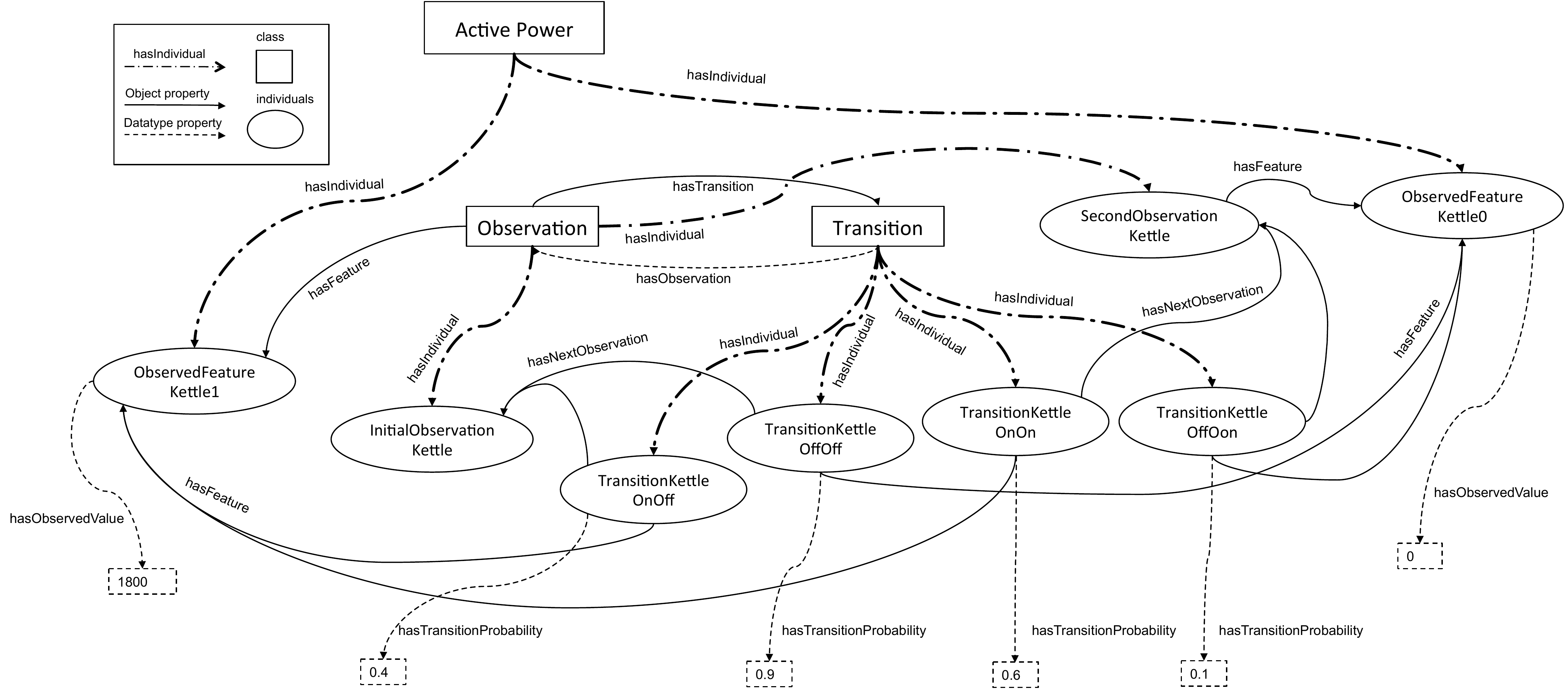}
\caption{Load identification model for the water kettle}
\label{fig:nilmmodelkettle}
\end{figure}

In order to model device profiles and device models, we used the the open source tool Prot\'eg\'e\footnote{http://protege.stanford.edu} to build models in the \ac{OWL}.
The resulting ontology is available for use at the MONERGY project webpage\footnote{http://www.monergy-project.eu/appliance-ontology/}.
Fig.~\ref{fig:datasheetkettle} shows an example profile for a water kettle.
The device is user driven and has a physical service to heat water.
The service demands 0.03 KWh and is currently in the OFF status.
The service takes place over one state, requiring 1800 W with 5\% tolerance being insensitive to interruption and start delay.
Fig.~\ref{fig:nilmmodelkettle} reports the load identification model for the water kettle.
To identify the device, this model describes OFF and ON observations, using active power as a feature.
As noticeable, device dynamics are captured using transition probabilities.

\section{Conclusions}\label{sec:conclusion}
This paper addressed the problem of device and data interoperability in HEMS.
In particular we analyzed building blocks of a \ac{HEMS} to identify main challenges.
We put ahead a general architecture for \ac{HEMS}, where requirements and characteristics are presented.
Although the current standardization effort will soon provide technologies to design smart appliances, many energy consuming and producing devices will be non-smart. Replacing all legacy devices by smart devices would be economically infeasible in many situations. Therefore, such legacy devices are assumed to have a significant impact on the energy consumption and thus, deserve consideration within energy management applications.
We discussed the application of load detection for the identification of running loads, as well as the integration of inferred information into HEM systems.
We advocate for a common description for smart and legacy devices, as it would offer a uniform interface to access features and data, with consequent complexity reduction for application developers.
A case study is provided to show the effectiveness of a load disaggregation algorithm on real data collected from a living environment.
We carried out two different approaches to monitoring and disaggregating appliances where the load disaggregator inputs the aggregated power profile of i) all appliances and ii) grouped devices.
A state-of-the-art load detection algorithm was applied to identify which monitoring approach is suitable to integrate legacy appliances into a \ac{HEMS}.
Our results demonstrate that each monitoring approach allows for the correct detection and therefore integration of appliances, according to the requirements defined.
We pointed out different advantages and disadvantages of these approaches, which should be considered during the design of the \ac{HEMS}.
We also showed how similarities between electrical devices affect the disaggregation process negatively.
Finally, we exercise how information from detected appliances can be annotated to be exchanged within a \ac{HEMS}.
To the best of our knowledge, this is the first paper showing how the integration of legacy devices into the \ac{HEMS} could take place, what requirements should be fulfilled and which limitations and challenges need to be considered.

\section{Acknowledgments}
This work was supported by Lakeside Labs GmbH, Klagenfurt, Austria and funding from the European Regional Development Fund and the Carinthian Economic Promotion Fund (KWF) under grant KWF-$20214\mid22935\mid24445$. 

%\section{Acknowledgements}
%The work of A. Monacchi, D. Egarter and W. Elmenreich is supported by Lakeside Labs, Klagenfurt, Austria and funded by the European Regional Development Fund (ERDF) and the Carinthian Economic Promotion Fund (KWF) under grant KWF 20214 | 23743 | 35470.

\bibliography{dominik,andrea}

\end{document}